\documentclass[%
 reprint,
 nofootinbib,
kgroupedaddress,
 amsmath,amssymb,
 aps,
]{revtex4-2}

\usepackage{booktabs}
\usepackage{todonotes}
\usepackage{marginnote}
\usepackage{multirow}
\usepackage{xcolor}
\usepackage{graphicx}
\usepackage{dcolumn}
\usepackage{bm}
\usepackage[colorlinks=true, allcolors=blue]{hyperref}

\begin{document}

\preprint{APS/123-QED}

\title{Invariant multiscale neural networks for data-scarce  scientific applications} 
\author{I.~Schurov$^1 {}^a$, D.~Alforov$^2$,  M.~Katsnelson$^1$, A.~Bagrov$^1$ and A.~Itin$^3$  }
 \affiliation{
     $^1$Radboud University, Nijmegen, The Netherlands\\
$^2$Independent\\
$^3$Hamburg University of Technology, Hamburg, Germany\\
$ {}^a $ \texttt{ilya@schurov.com}
}

\date{\today}

\begin{abstract}
Success of machine learning (ML) in the modern world is largely determined  by abundance of data. However at many industrial and scientific problems, amount of data is limited.  
Application of ML methods to data-scarce scientific problems can be made more effective via several routes, one of them is equivariant neural networks possessing knowledge of symmetries. 
Here we suggest that combination of symmetry-aware invariant architectures and stacks of dilated convolutions is a very effective and easy to implement receipt allowing
sizable improvements in accuracy over standard approaches. We apply it to  representative physical problems from different realms: prediction of bandgaps of photonic crystals, and network approximations of magnetic ground states.  The suggested invariant multiscale architectures increase expressibility of networks, which allow them to perform better in all considered cases.

\end{abstract}

\maketitle

\section{Introduction}

ML methods have seen various applications in science \cite{PhysicsML}. 
It has been suggested that we are entering new, fifth paradigm of scientific discovery \cite{Fifth} (the forth paradigm being related to data-driven approaches to understanding large volumes of experimental data \cite{FourParadigm}, while in the fifth paradigm  data  comes not from experimental observations, but from numerical simulations of the fundamental equations of science). 
Datasets  obtained via solver calculations are often very different from those encountered in traditional ML applications: they are super-clean and respect well-defined physical laws. At the same time, the amount of available data is often scarce as it is expensive to produce. It is therefore important and useful to use suitably adapted ML methods that respect physical symmetries and other properties of data. 
Another major complication in scientific ML as compared to the industry is as follows. 
The use of industrial large-scale neural networks requires computational resources and highly specific expertise that are usually lacking in scientific research communities. Deploying and training a network of tens of millions or billions parameters is not always a feasible task to perform as a part of a fundamental research project in physics or chemistry. Another side of the coin is that the same research group can switch between conceptually different problems, and designing tailored architectures and learning algorithms for each task can be unbearably time-consuming. Hence, it is of critical importance to suggest ML algorithms that are powerful and versatile enough to address diverse problems in different areas of natural sciences with only marginal alterations, and, simultaneously, compact and simple to deploy to be routinely used by domain experts without extensive experience in ML. In this paper, we attempt to address both issues and suggest a simple and light neural network architecture that can be leveraged to solve problems in classical and quantum physics with limited in size and low-dimensional datasets.

It is based on a combination of translationally invariant Convolutional Neural Network (CNN) architecture and stacks of dilated convolutions. We extend approach of \cite{Invariant}, where translationally invariant CNN-based architectures were studied, adding ordered sequences of dilated convolutions and considering photonics \cite{ML1,MIT1,MIT2,Boltasseva2021,Sigmund2008} and quantum physics applications \cite{NQS,NQS1,NQS2,NQSB}.

In photonics realm, we are interested in photonic crystals \cite{PhBook,Sigmund2008}:  optical nanostructures where refractive index changes periodically in space in one, two, or all three dimensions. This affects propagation of light and have (or may have) important technological applications in  photovoltaic cells, mirrors, sensors, fiber-optic communication, and photonics computers. Photonic crystals are not only manufactured industrially, but are encountered in nature: e.g. in wings of butterflies \cite{wing1, wing2} and precious opals \cite{opal} (see also \cite{Tailored} and references therein). 
Designing artificial photonic crystals for specific applications using computer simulations is often very demanding problem. Usage of machine learning methods was suggested \cite{ML1,MIT1,MIT2} to improve this process.

Another important domain of scientific applications of neural networks is many-body quantum physics. In the framework of quantum mechanics, a system of interacting electrons or localized magnetic moments on a crystalline lattice is described by its Hamiltonian matrix $\hat{H}$, which linear dimension scales exponentially with the number of particles. Obtaining spectrum of this matrix is synonymous to having a complete solution to the quantum problem of interest. In reality, due to the exponential size of the matrix, this problem is unfeasible in most of the cases. Even finding the so called ground state $|\psi_0 \rangle$ -- the eigenvector with the lowest eigenstate -- representing wave function of the system at zero temperature is an important problem, which solution can shed light on many aspects of phenomenology of interacting quantum particles:
\begin{equation}
    \hat{H}|\psi_0 \rangle= E_0 |\psi_0 \rangle, \,\,\, E_0 = \min \mbox{Spec}(\hat{H}).
\end{equation}
Since $\dim(|\psi_0 \rangle) \sim 2^N$ for spin-1/2 localized magnetic moments and $\dim(|\psi_0 \rangle) \sim 4^N$ for itinerant electrons in a single band, this problem can be solved exactly only for a handful of degrees of freedom. For a larger number of constituents (ideally one should be able to describe a few hundred particles), one should resort to approximate methods. In this context, machine learning can be of a great use as a variational approach to finding the ground state. As we will explain in detail in the corresponding section, in many cases, the target vector $|\psi_0 \rangle$ can be represented as a signful real-valued variational function of binary variables. Given that this variational ansatz is expressive enough and has high generalization capacity, by optimizing its parameters, one should be able to represent the ground state with good accuracy. Using a neural network as such variational ansatz, and identifying its loss function with energy of the state, the ground state can be obtained by means of neural network learning. This paradigm is known as neural-network quantum states (NQS)~\cite{NQS}. In this paper, we will demonstrate how dilated CNN can be used to improve accuracy of NQS for so called frustrated quantum systems, where competing interactions cause the structure of the ground state to be complex and difficult to represent.

The aforementioned systems are often characterized by a number of different length scales. For example, many-body quantum systems can exhibit so called quantum critical points and be (nearly) scale-invariant in their vicinity. Hence, traditional neural networks focusing on local structures, like CNN with conventionally defined local filters, can be not an optimal choice for analyzing such systems. We will show that dilated convolutions indeed help a lot when dealing with multi-scale physical systems.

The paper is organized as follows. In Sec. \ref{sec:architectures}, we define the dilated CNN and describe their architecture. In Sec. \ref{sec:photonics}, the dilated CNN are used to predict band structures of photonic crystals and shown to outperform standard architectures. In Sec. \ref{sec:NQS}, networks of this class are successfully employed to learn sign structures of many-body quantum ground states. We conclude with an outline of possible applications that go beyond considered in this paper.

\section{Translationally invariant CNN architectures and multiscale (dilated) convolutions}
\label{sec:architectures}
\subsection{Translationally invariant CNN architectures}
\begin{figure}
\caption{ (a) Top: Typical CNN architecture with only approximate translational invariance (See also Ref.~\cite{Invariant}) 
(b) Bottom: CNN architecture with exact translational invariance
 }
\label{Fig1a}
\centering
\includegraphics[width=79mm]{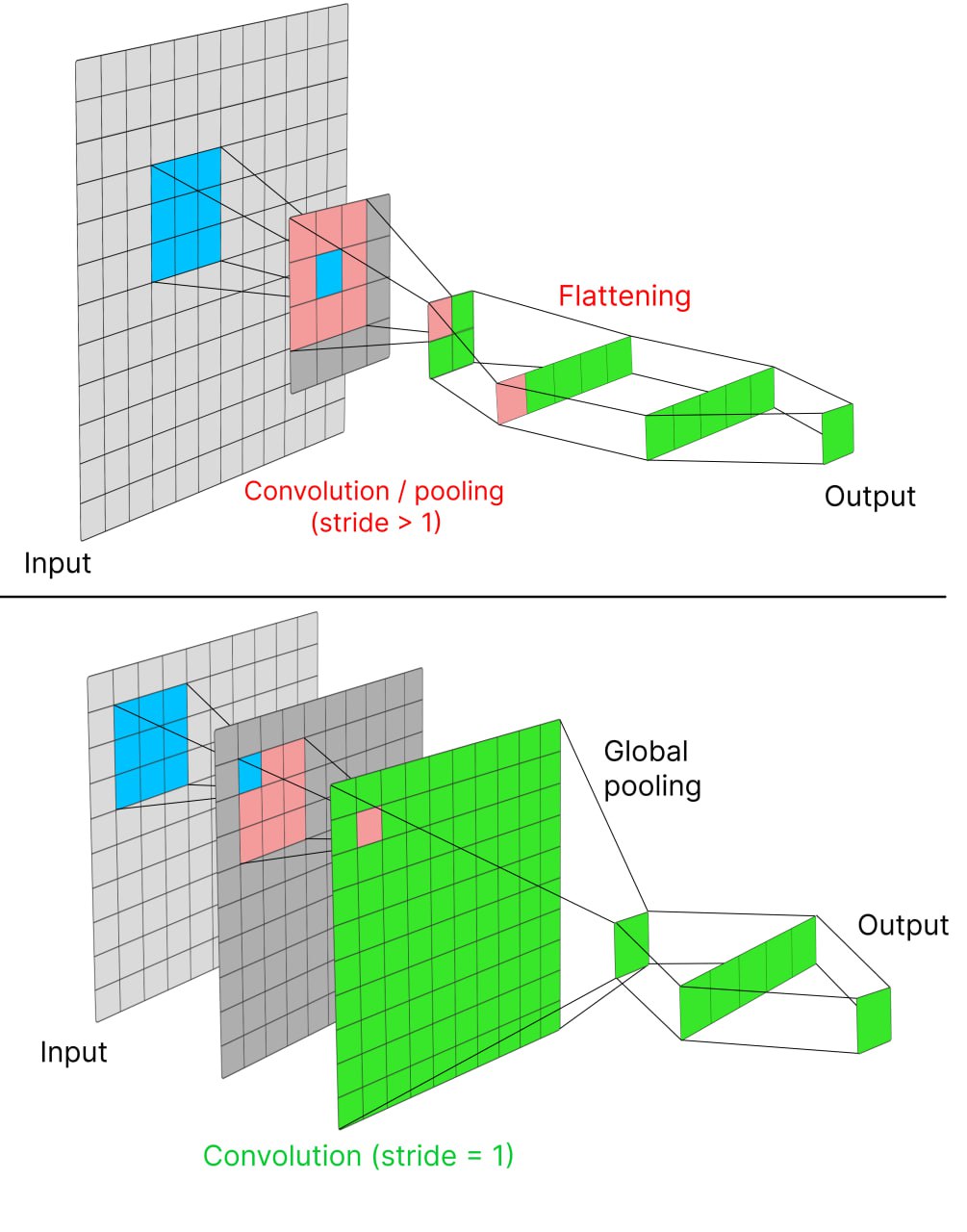}

\end{figure}

Typical CNN architectures (Fig.~\ref{Fig1a}a) involves several convolutional layers consisting usually of kernels (filters) of small or moderate size, batch normalisation operations \cite{ioffebatch}, nonlinear activation functions,  max pooling layers, etc.

\begin{figure}
\caption{The concept of dilated convolutions. Dark squares designate non-zero parameters of a convolutional kernel. At light squares, parameters are zero. Even though the size of the kernel grows quadratically with the dilation parameter, the number of parameters remains the same. Left: a convolutional kernel of size $3\times 3$ without dilation (or, equivalently, with dilation parameter equal to 1). Right: the same $3\times 3$ convolutional kernel with dilation=2.
\label{Fig2}}
\centering
\includegraphics[width=80mm]{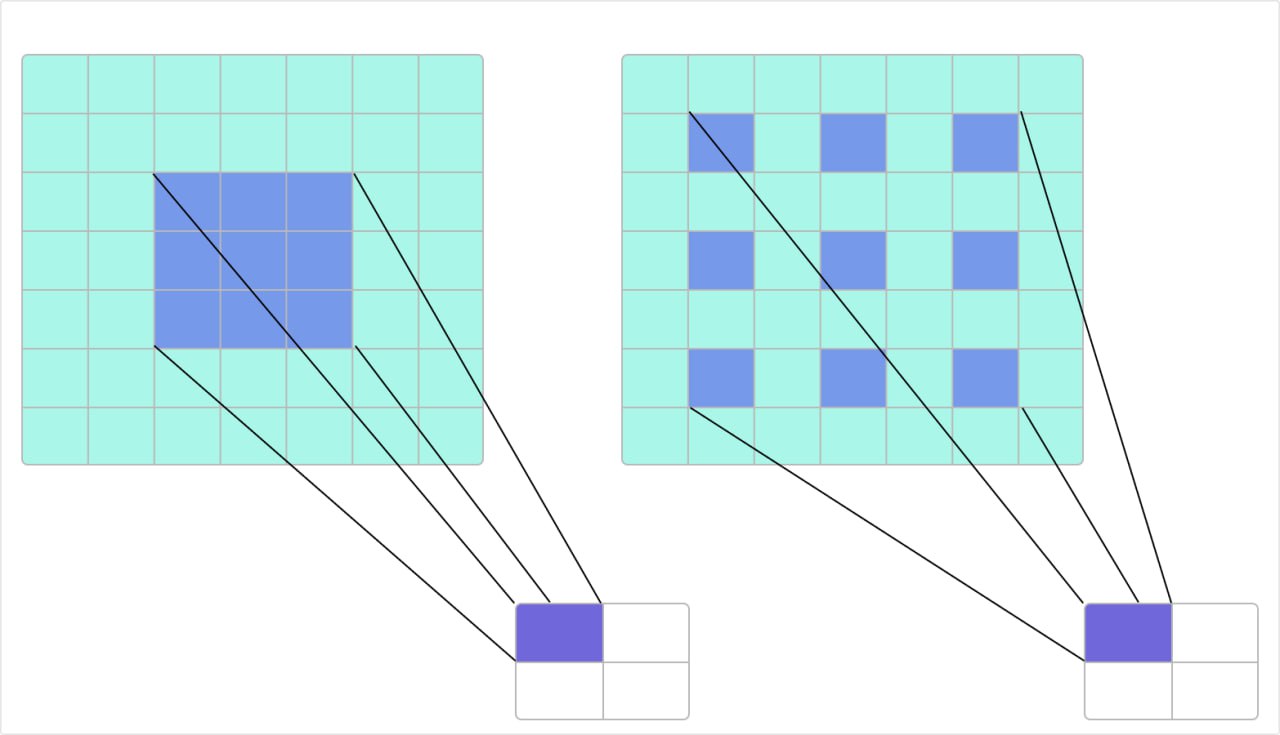}

\end{figure}

They often follow the same general design principle of successively applying convolutional layers to the input, consequently downsampling the spatial dimensions while increasing the number of feature maps (in order to keep complexity of passing a layer approximately constant).
While for general ML applications (such as computer vision tasks) CNN is often considered to be a translationally invariant network, it is easy to check that this invariance is in fact only approximate. A cyclic shift of image on just one pixel would result in change in the output. As explained in detail e.g. in \cite{Invariant},  spatial pooling operations, and convolutions with stride greater than one destroy translational invariance. Also, between convolutional part and fully connected part of the network there should be no flattening layer; instead, global pooling operation can be applied in order to achieve exact translational variance. As a result, translationally invariant architecture is pretty much different from traditional designs. There are no downsamplings along the network, and the last convolutional layer is ended with global pooling operation (Fig.~\ref{Fig1a}b). Intuitively, in order to enhance analysis on larger spatial scales, we need to increase size of the kernels from layer to layer (because usually size of feature maps is decreased from layer to layer and convolutional kernels of the same size effectively operate on larger and larger parts of the image). However, kernels of large size are cost-ineffective. Dilated convolutions, having been introduced in \cite{Holschneider}, offer an attractive solution for this problem increasing kernel sizes without increasing number of parameters \cite{Chen2017, medium} (Fig.~\ref{Fig2}). 

\subsection{Dilated convolutions and stacks of dilated convolutions}
 Compared to
the usual convolution, the dilated one
has a hyper-parameter "dilation rate", which refers to
the number of intervals between non zero elements of the receptive field (i.e. normal convolution has dilatation rate equal to 1). In one dimension, dilated convolution
can be defined as \cite{Wang18, Ku21}
$$ g[i] = \sum_{l=1}^L f[i + d \cdot l] \cdot w[l],
$$
where $f[i]$ is the input signal, $g[i]$ is the output signal, $w[l]$  denotes the filter of length $L$, and $d$ corresponds to the dilation rate of the filter. In usual convolution, $d = 1$.

Stacks of dilated convolutions with consequently increasing dilation were used e.g. in Wavenet architecture \cite{Wavenet, Wavenet2} for audio applications, and in many computer vision projects \cite{dilexamples}.
However, combination of the translationally invariant architecture and stacks of dilated convolutions is not usual: in industrial computer vision applications one usually is not concerned with exact translational invariance of the network, whereas in scientific applications where translational invariance is ensured \cite{Invariant}, addition of stacks of dilated convolutions is not usually done.

\section{Experiments with photonics datasets}
\label{sec:photonics}

\begin{table*}[t]
\begin{tabular}{ |p{3cm}||p{3cm}|p{3cm}|p{3cm}| }
 \hline
 \multicolumn{4}{|c|}{Experiments with the $D_3$ dataset} \\
 \hline
  & Model $a$, with dilations  & Model $b$, no dilations & Model $c$, no translational invariance  \\
 \hline
 Dilations & $\{1,2,4,8,16\}$ & $\{1,1,1,1,1\}$ & $\{1,1,1,1,1\}$   \\
  Strides & $\{1,1,1,1,1\}$ & $\{1,1,1,1,1\}$ & $\{1,2,1,2,1\}$   \\
 Channels sequence   & $\{24,24,24,24,24\}$    & $\{24,24,24,24,24\}$ & $\{24,24,24,24,24\}$  \\
 Nr. of parameters  &  46,8K   & 46,8K   &  46,8K          \\
  \hline
    \multicolumn{4}{|c|}{MAPE results, averaged over 4 folds and 4 launches}  \\
    \hline 
 On ``the best epoch''  & {\bf 0.83}   & 1.14   &  0.96            \\   
 On ``the last epoch''&  {\bf 1.25 }   & 1.56 &  1.43  \\ 
 \hline
\end{tabular}
 \caption{Experiments with the $D_3$ dataset. Cross-validation over 4 folds was fulfilled, with 4 random launches.}
  \label{tab:D3.2}
\end{table*}

\begin{table}[t]
\begin{tabular}{ |p{3cm}||p{3cm}| }
 \hline
 \multicolumn{2}{|c|}{Optuna experiments with the $D_3$ dataset} \\
 \hline
 Parameter &        \\
 \hline
 Number of training samples & 800      \\
 Channels sequence   & $\{16,16,16,16,16\}$         \\
 Variants of dilation orderings&   [$V_1,V_2, V_3$]         \\
 Nr of trials & 48       \\
 Nr. of variants     & 96       \\
  \hline
   
    \multicolumn{2}{|c|}{Results of the best trial, averaged over 4 launches}  \\
    \hline 
 CFactor& 2           \\
 Number of parameters  &  55.1k                     \\
 Dilation variant & $V_2$          \\
 Filter size, 1st layer& 3        \\
 Filter size, the last layer& 1         \\
 MAPE & 0.67           \\
 \hline
\end{tabular}
 \caption{Optuna experiments with the $D_3$ dataset. 48 trials were done, every trial was launched 4 times for averaging.  Number of convolutional layers: 5, number of FC layers: 3. 3 variants of ordering dilations were considered: $ V_1: \{1,2,3,6,12\}$, $ V_2: \{1,2,4,8,16\}$, $ V_3: \{1,1,1,1,1\}$. CFactor was chosen from the set $[1,2]$. Number of channels in the convolutional part were fixed as $\mathrm{CFactor} \times \{16,16,16,16,16\}$. Convolution filter sizes in the first and the last layers were chosen from the set $[1,2,3,4]$. }
  \label{tab:D3.3}
\end{table}

We consider dielectric materials periodically distributed in $xy$-plane and constant in $z$- direction, i.e. $\epsilon( \bf{x} + \bf{R_j}) = \epsilon(\bf{x})$ , where $\bf{R_j}$ are lattice vectors with zero $z$- component, $\epsilon$ is the dielectric permittivity. Moreover, the dielectric function $\epsilon$ is piece-wise-constant and only attains two possible values: $ \epsilon_1, \epsilon_2$.
The Maxwell equations for electromagnetic waves propagating in $xy$-plane are decoupled into two wave equations for TM 
(transverse magnetic, with $\bf{E}$-field in the $z$- direction) and TE (transverse electric, with $\bf{H}$-field in $z$- direction) polarized waves:
\begin{equation}
TM : \nabla^2 E_z ({\bf{x}}) + \frac{\omega^2 \epsilon({\bf{x})}}{c^2} E_z ({\bf x}) =0, \label{TM}
\end{equation}

\begin{equation}
TE : \nabla \cdot \Bigl[ \frac{1}{\epsilon ({\bf x}) } \nabla H_z ({ \bf x}) \Bigr] +\frac{\omega^2 }{c^2} H_z ({\bf x}) =0.
\end{equation}

\begin{figure}
\caption{Dataset of photonic crystals.  Top row:  an example of a unit cell consisting of two materials with permittivities $\epsilon_{1,2}$ and a set of such unit cells. Center: a unit cell generates a periodic 2D structure (photonic crystal). Bottom row: eigenstates of Maxwell equations are characterised by two numbers (quasimomenta $k_{x,y}$) being located in the first Brillouin zone. A $23 \times 23$ grid was defined, and at each point the first 6 eigenstates were calculated using MPB solver.  \label{Fig3}}
\centering
\includegraphics[width=80mm]{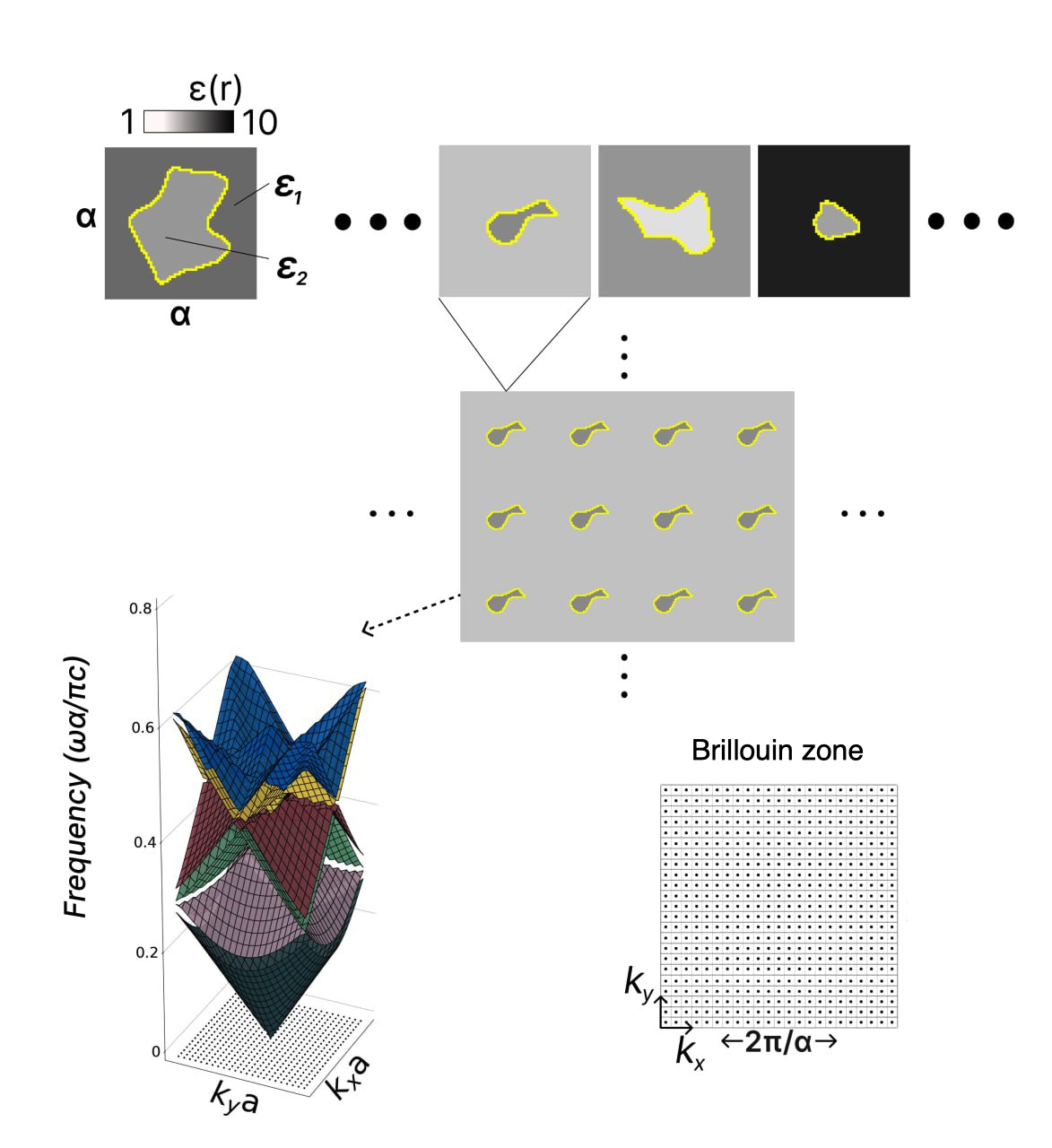}

\end{figure}

The scalar fields satisfy the Floquet-Bloch wave conditions $E_z = e^{i {\bf k \cdot x}} E_k$ and $H_z = e^{i {\bf k \cdot x}}H_k$, respectively, where $E_k$ and $H_k$ are cell periodic fields \cite{Sigmund2008}.
Concentrating on \eqref{TM} for definiteness, and solving it for wave numbers $k$ belonging to the first Brillouin zone we get a band diagram as shown in Fig.~\ref{Fig3}.  
A complete bandgap between bands $n$ and $n+1$ exists if
$\min_k \omega_{n+1} > \max_k \omega_n$.
Usually one defines a relative bandgap as the ratio of the absolute bandgap to the midgap frequency:

\begin{equation} \Delta \omega_n = 2 \frac{\min_k \omega_{n+1} - \max_k \omega_n}{\min_k \omega_{n+1} + \max_k \omega_{n}}
\end{equation}

A typical task of photonic crystal design is to create a structure with the largest possible bandgap. 
A neural network which can predict band diagram or band gaps of the photonic crystal from the unit cell is very useful for this purpose \cite{MIT1,MIT2,ML1}.
We made experiments with three datasets of photonic crystals ($D_1, D_2, D_3$). $D_1$ dataset is the same as was used in \cite{MIT1,ML1}. It contains images of $64\times 64$ size, each of them  defining a unit cell of a two-dimensional photonic crystal. Values of pixels designate optical permittivities $\epsilon_1, \epsilon_2$ of the two materials,  that are randomly distributed in the interval  $[1,10]$ (see Fig.~\ref{Fig3}).
These values roughly correspond to permittivities of transparent materials in the visible spectrum (e.g. at  wavelengths of 500-700 nm, the permittivity of air, silica (silicon dioxide), silicon nitride, and silicon carbide is approximately 1.0, 2.1, 4.0--4.1, and 6.9--7.2, respectively). 
The dataset contains band diagrams of the first 6 bands calculated using a physical solver MPB \cite{mpb}. 

\begin{figure}
\caption{Examples of $96 \times 96$ unit cells from the third dataset. Complexity factor F=1,2,3,4,5,6 from top left to the bottom right.   \label{FigSt}}
\centering
\includegraphics[width=70mm]{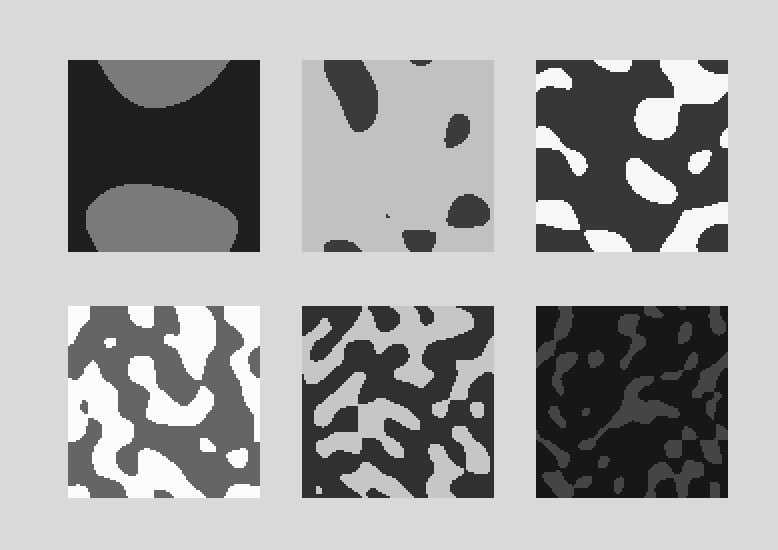}

\end{figure}

Detailed description of the other two datasets, $D_2$ and $D_3$, are available in Appendix B. They contain more complex structures (see Fig.~\ref{FigSt}) and more bands are calculated for each sample (17 bands instead of 6).

In the experiments, we train neural networks to predict $2N-1$ values for each unit cell sample:  maxima and minima of $N$ bands (except the minimum of the first band which is always 0), where $N=6$ in the $D_1$ dataset and $N=17$ in  $D_2,D_3$ datasets. These values determine appearance and disappearance of bandgaps and clearly have advantage over predicting bandgaps themselves: the latter are usually zero for most of the structures, whereas the former have certain nonzero values for any structure.  
As a metric, we use Mean Average Percentage Error (MAPE).

Experiments with the $D_3$ dataset (containing 1000 images of $96 \times 96$ resolution, Fig.~\ref{FigSt}) are shown in  Tables \ref{tab:D3.2}, \ref{tab:D3.3}. 

Here we predicted maxima and minima of the 17 bands, i.e. 33 numbers.
We consider a set of particular models (models $ a,b,c$ in Table~\ref{tab:D3.2}). They are non-optimised and are just examples of reasonably prepared architectures.
Models $a$ and $b$ differ only by the dilation block (in the model $a$ it is \{1,2,4,8,16\}, while in the model $b$ it is \{1,1,1,1,1\}, i.e no dilations at all). The model $c$ is built in a ``traditional'' way, with convolutional layers of stride 2 and therefore with downsampling and loss of translational invariance. For each model we fulfilled 4 launches of 4-fold cross-validation and averaged the result (i.e., each model was launched 16 times).  We calculated the result in two different ways. In the first way, we determine the best epoch on validation, and infer the model at this particular epoch on a test dataset. That is, during cross validation procedure the dataset is split into $K=4$ ``folds'', and three folds are used for training, while the fourth fold is further split into the evaluation and test datasets. During 200 epochs of training, a model is being evaluated on the evaluation dataset, and its state during the best performing epoch is saved. In the end, performance of this saved model on the test dataset is  recorded as its result ``at the best epoch''. In the second way, we just test a trained model after 200 epochs of training. In real applications, usually a specialised  algorithm is used for the scheduling of training, e.g. early stopping algorithm. It introduces additional hyperparameters of the model architecture which can be adjusted (``patience'', etc), and the result is often sensitive to these hyperparameters. For our purposes, the two results mentioned above provide reliable estimate of models performance. 

From results of the simple experiments given in Table~\ref{tab:D3.2}, it follows that an invariant architecture introduced in replacement of the traditional downsampling architecture may lead to slight deterioration of accuracy, if not accompanied by neural architecture search (NAS), which we have done using the Optuna package~\cite{optuna} (see Tables \ref{tab:D3.3}, \ref{tab:D1.0}, \ref{tab:D1.01}). 
However, stacks of dilated convolutions generally improve results of the bare invariant architecture considerably, so the model $a$ performs better than both invariant architectures without dilations ($b$) and downsampling architectures with loss of invariance (i.e, $c$). In other words, stacks of dilated convolutions seem to be a natural receipt replacing downsampling layers in usual CNNs. Similar experiments of this type (not shown) seem to confirm such behaviour.
To check this idea more systematically,
in Table~\ref{tab:D3.3} we present results of NAS optuna \cite{optuna} experiments with the $D_3$ dataset, where the selected optimised architecture is the one with stacks of dilated convolutions. 
This was confirmed for $D_2$ dataset as well (not shown),
and with experiments with $D_1$ dataset being presented in Appendix B.
Dataset $D_1$ contains many samples (20k), but their structure is not as complex as in $D_2$ and $D_3$.
Unlike approach of \cite{MIT1,ML1}, where all the data was used to train a network, in the experiments with the $D_1$ dataset we consider training of models on sub-datasets of variable sizes $N_{train}$, following approaches of \cite{MIT2}. 

In the first set of experiments with $D_1$ dataset (Table~\ref{tab:D1.0}) 
we launch 4 sessions of experiments where optimal architecture is chosen from
prescribed configuration space. E.g., in Session 1 there were 128 possible network architectures. Ordering of dilations in convolutional layers were either $\{16,8,4,2,1\}$ (variant $V_1$), or $\{1,2,4,8,16\}$ ($V_2$), or $\{1,2,4,8,12\}$ ($V_3$), or $\{1,1,1,1,1\}$ ($V_4$, no dilations). Number of trials were 64, i.e. 64 architectures were tried out of the set of 128. MAPE of the found optimised architecture was 0.51, and its dilation variant was $V_1$ (however suboptimal architectures with dilation variants $V_2$, $V_3$ had similar values of MAPE, whereas the best architecture with $V_4$ variant (no dilation) in the Session 1 had MAPE $>$ 0.7. One can conclude that stacks of dilated convolutions produce reliable improvement, but no preferable ordering of these convolutions can be determined yet). For comparison, in Sessions 2-4, where no dilations were used ($V_4$), similar hyperparameter optimisation process produced result with MAPE larger than 0.7, which again suggests usefulness of dilations.   All experiments were done with $N_{train}$=1600 training samples. Network architectures contain 5 convolutional layers and 3 fully connected layers.  Number of channels in subsequent convolutional layers were chosen according to $\{1,1,1,1,1\} \times \mathrm{CFactor}$ (Sessions 1,2) or $ \{1,1,2,2,4\} \times \mathrm{CFactor}$ (Sessions 3,4), where the hyperparameter CFactor was chosen from the configuration set $[24,48]$ in the former case and $[12,24]$ in the latter, in order to approximately equalize number of parameters in the architectures under consideration. Configuration space of architectures in Session 1 and Session 2 was exactly the same, apart from dilation variants.  Session 4 contained strided convolutions, i.e. non invariant architectures. Training was performed for 200 epochs. We chose the result at the best epoch on validation as the final result. The fact that optimized architectures without dilations in all four sessions of experiments performed worser than optimised dilated variants of Session 1 seem to be convincing argument in favor of dilations.

Additional experiments with the first dataset are reported in  Table~\ref{tab:D1.01}. We choose $N_{train}=800$ and fixed CFactor in each session, but varied configuration of the fully connected part of network. Similar conclusions can be obtained from this table as well.  
Batch normalisation was not used in these experiments, from a separate set of experiments (not shown) we found it does not improve the results.
Typical improvements of accuracy for usage of dilations are about 15--20\%. 
This is already a sizable and useful result. E.g., from somehow related calculations of \cite{MIT2} it could be seen that for improving accuracy on $\sim$ 30\% one usually needs an order of magnitude more data. To achieve 20\% improvement of accuracy, one needs much more than 20\% increase of amount of data (say, two-fold increase or even more). Moreover, we expect that for datasets of more complicated structures improvements would be even higher. Indeed, experiments with the third dataset being reported in Tables~\ref{tab:D3.2}, \ref{tab:D3.3} seem to support this conjecture, although more systematic experiments would be needed.




\section{Experiments with neural quantum states}
\label{sec:NQS}
\begin{figure}
    \centering
    \includegraphics[width=0.5\textwidth]{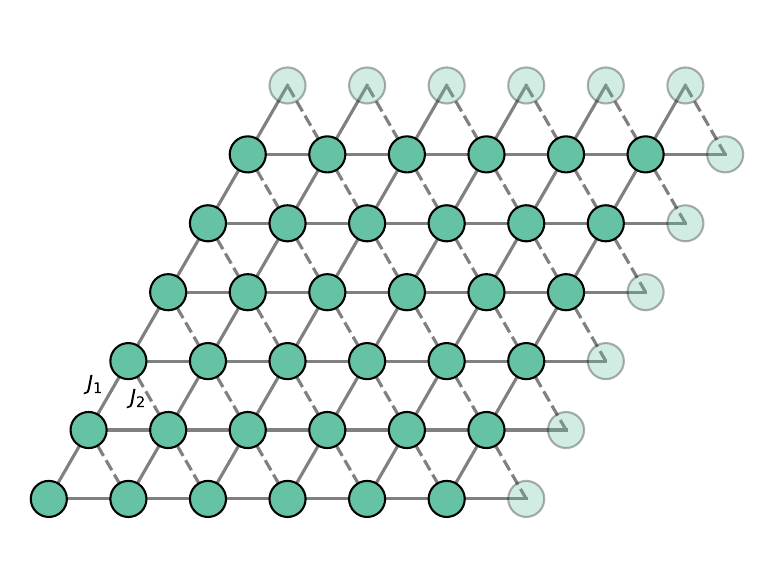}
    \caption{Triangular 36-spin lattice with periodic boundary conditions. Different couplings $J_1$ and $J_2$ are taken along different directions, with $J_2=0$ corresponding to a bipartite (square) lattice with simple ground state structure, and $J_2/J_1>1.25$  -- to strongly frustrated regime with spin liquid ground state.}
    \label{fig:triangular-lattice}
\end{figure}

As a playground for quantum mechanical applications, we shall consider quantum spin-1/2 Heisenberg antiferromagnet on an $N$-site triangular lattice (see Fig.~\ref{fig:triangular-lattice}) with two different coupling constants described by the Hamiltonian:
\begin{gather}
    \hat{H} = J_1\sum\limits_{\langle k,l\rangle}\left( \sigma^x_k\otimes \sigma^x_l + \sigma^y_k\otimes \sigma^y_l + \sigma^z_k\otimes \sigma^z_l\right)+ \label{eq:Heisenberg} \\
    J_2\sum\limits_{\langle\langle m,n\rangle\rangle}\left( \sigma^x_m\otimes \sigma^x_n + \sigma^y_m\otimes \sigma^y_n + \sigma^z_m\otimes \sigma^z_n\right) \nonumber
\end{gather}
where $J>0$, $\sigma_k^{x,y,z}$ are Pauli operators acting on site $k$, and the summation goes over the lattice bonds. The $\otimes$ symbol here represents a short-hand notation for a Kronecker product of a sequence of $N-2$ identity matrices standing in positions $\neq k, l$, and two Pauli matrices in positions $k$ and $l$, so that each term in \eqref{eq:Heisenberg} is a $2^N-\mbox{by}-2^N$ matrix. We will be interested in the ground state of this system -- a state with the minimal energy (eigenvalue of \eqref{eq:Heisenberg}).
Any eigenstate wave function of such a system can be decomposed as a linear combination of the Hilbert space~$\cal H$ basis vectors
\begin{equation}
    |\psi\rangle = \sum\limits_{i=1}^{2^N} \psi_i |i\rangle, \label{eq:expansion}
\end{equation}
with $\psi_i \in \mathbb{R}$ \footnote{In generic case, coefficients of the wave function are complex, but for a large class of quantum systems including \eqref{eq:Heisenberg}, the eigenstates are real-valued.}, and $\sum\limits_i |\psi_i|^2=1$. We will be using two representations of basis vectors $|i \rangle$. On the one hand, each basis vector corresponds to a classical spin configuration of the system with spins ``up'' encoded as $1$ and spins ``down'' encoded as $0$: $|i \rangle = | 1, 0, 0, \dots 1, 0\rangle$, and can be viewed as a binary sequence of length $N$. On the other hand, $|i\rangle$ are spanning the space of states of the exponentially large matrix \eqref{eq:Heisenberg}, and, in this context, are represented as vectors of length $2^N$.

In the variational approach, coefficients of \eqref{eq:expansion} are represented as values of parametrized function:
\begin{equation}
    \psi^{\bf W}_i = \psi({\bf W}, |i \rangle): \{0,1\}^N \rightarrow \mathbb{R}, \label{eq:variational}
\end{equation}
where $\bf W$ is the set of variational parameters. In particular, one can represent the wave function as a neural network -- NQS -- which takes binary sequences of spins as inputs and returns wave function coefficient as outputs. $\bf W$ will then be weights of the network, and the energy of the state $E=\langle \psi^{\bf W} |\hat{H}| \psi^{\bf W}\rangle$ plays the role of loss function, by minimizing which one can approximate the ground state.

One critical aspect of learning must be highlighted here. Formally speaking, energy of the quantum state is computed by summing over the whole Hilbert space basis, which is impossible for a system of more than $\sim 40$ particles:
\begin{equation} \label{eq:energy}
    E=\langle \psi^{\bf W} |\hat{H}| \psi^{\bf W}\rangle = \sum\limits_{i,j=1}^{2^N} \langle i | \hat{H} | j \rangle \psi_i^* \psi_j. 
\end{equation}
Hence, it must be approximated. This is normally done by means of Markov Chain Monte Carlo (MCMC) sampling. Rewriting \eqref{eq:energy} as
\begin{gather}
 E = \sum\limits_{i=1}^{2^N} \langle \psi^{\bf W} | i \rangle \langle i | \hat{H} | \psi^{\bf W} \rangle = \sum\limits_{i=1}^{2^N} \frac{\langle i | \hat{H} | \psi^{\bf W} \rangle}{\langle i | \psi^{\bf W} \rangle} |\psi_i|^2 = \\ 
 = \sum\limits_{i=1}^{2^N} E_{\mbox{loc}}(i) p(i), \nonumber
\end{gather}
one can see that energy of a quantum state can be represented as an expectation value of a function dubbed $E_{\mbox{loc}}(i)$ with respect to non-negative probability distribution $p(i) = |\psi_i|^2$ defined by the coefficients of the wave function, which is a conventional concept in classical probability theory. Such an expectation value can be estimated by sampling a finite subset of $K$ basis vectors and approximating $E$ with a much shorter sum:
\begin{equation}
    E = C\cdot \sum\limits_{n=1}^{K} E_{\mbox{loc}}(i_n) p(i_n), 
\end{equation}
where $C$ is a normalization constant equal to $(\sum_{n=1}^K p(i_n))^{-1}$ and $K$ is usually taken to be about $2000\cdot N$.

This approximation requires variational ansatz \eqref{eq:variational} to have good generalization capacity. At every optimization step weight $\bf W$, and hence the wave function coefficients $\psi^{\bf W}_i$, are updated based on the values of energy $E$ and its gradient $\partial_{\bf W} E$ evaluated on a small ($K$-size) subset of the Hilbert space basis. In order to be globally correct, the ansatz should correctly predict coefficients $\psi^{\bf W}_i$ of vectors $|i \rangle$ that have not been encountered during the MCMC sampling. For a real-valued wave function, the coefficients can be decomposed in the absolute value and sign parts as:
\begin{equation}
    |\psi\rangle = \sum\limits_{i=1}^{2^N} {\cal S}_i|\psi_i| |i\rangle,\,\, {\cal S}_i \in \{-1,+1\}.
\end{equation}
The set of all signs ${\cal S}_i$ is known as the sign structure, and two different neural networks can be used to represent the absolute values and the sign structure. In~\cite{NQSB}, it was shown that one of the critical problems with using neural quantum states is their insufficient generalization capacity when learning the sign structures of the frustrated quantum systems ground states. At the same time, learning absolute values represents a challenging but solvable problem. Hence, in this paper we focus specifically on learning the sign structure \cite{CommPhysWesterhout}.

For that, we follow \cite{NQSB}, and consider a setup of supervised sign structure learning, which is different from energy optimization. The following steps are taken:
\begin{itemize}
    \item For each model of interest, we perform exact diagonalization of a system of up to 36 spins using \texttt{lattice-symmetries} package~\cite{Westerhout2021,Westerhout2023} and obtain ground state of the model:
    \begin{equation}
        |\psi_0 \rangle = \sum\limits_{i=1}^{2^N} {\cal S}^{0}_i|\psi^{0}_i| |i\rangle.
    \end{equation}
    This provides us with data to conduct supervised learning and with ground truth to assess the quality of learning (superscript $0$ denotes the ground truth signs and absolute values).
    \item We sample a limited number (different in different experiments) of elements from probability distribution $p(i) = |\psi^{0}_i|^2$ to construct training set $\cal C$ of basis vectors, and $10000$ basis vectors to construct test set $\tilde{\cal C}$. Training and test sets are ensured not to have common elements.
    \item A neural network architecture is picked to define a map from basis vectors (binary sequences of classical spin configurations) to corresponding signs:
    \begin{equation}
        {\cal S}^{neural}: | i \rangle \rightarrow \{ -1, +1\}.
    \end{equation}
    Computationally, the output of the neural network is an estimate of the probability that the sign is positive (denoted by ${\cal P}^{neural}$), that is later thresholded at the value of $0.5$ to obtain the actual prediction, i.e. ${\cal S}^{neural}_i$ is equal to $+1$ if ${\cal P}^{neural}_{i}>0.5$ and $-1$ otherwise.
    \item We use the knowledge of exact absolute values $|\psi_i|$ and signs ${\cal S}_i^{0}$ to train the neural network to minimize the loss defined as the binary cross-entropy:
    \begin{align*}
        BCE = -\sum\limits_{|i\rangle \in {\cal C}} (&[S_i^0=+1]\log {\cal P}^{neural}_i\\
        +& [S_i^0=-1]\log (1-{\cal P}^{neural}_{i})),
    \end{align*}
    where square brackets denote a value that is equal to $1$ if condition is satisfied and $0$ otherwise.
    \item Quality of learning is measured by the so-called \emph{sign overlap} on the test set:
    \begin{equation}
        \tilde{\cal O}_{\cal S} = \sum\limits_{|i\rangle \in \tilde{\cal C}} |\psi^{0}_i|^2 {\cal S}_i^{0} {\cal S}_i^{neural}
    \end{equation}
\end{itemize}
Sign overlap can be considered as a rescaled and weighted accuracy, where weights are the squares of the amplitudes of the true ground state. If sign structure is learned and predicted exactly, $|{\cal O}| = 1$ (in quantum mechanics, global phase rotation does not change observable properties of the state: $|\psi_0 \rangle \rightarrow e^{i\varphi}|\psi_0 \rangle $. Hence, only absolute value of $\cal O$ matters, and ${\cal O} = -1$, with all signs flipped, would mean an exact solution).

It is known that ground states of the considered Hamiltonians lie in a specific sector of the full Hilbert space: their total magnetization should be zero, i.e. they only involve basis vectors $|i\rangle$ with equal number of $0$ and $1$ entries. 
We take small subsets of basis vectors belonging to this sector as our training and test sets. The size of the train set is equal to the size of the full sector times $\varepsilon_{train} \ll 1$ for specific small values of $\varepsilon_{train}$. The size of the test set is always $10000$ in our experiments. 

For the considered system, the ground state is invariant under the action of lattice symmetries, making it natural to use invariant networks. We choose the training and test sets in such a way that trajectories of the training set elements under the action of the lattice symmetry group do not intersect the test set, making sure that there is no leak of the test data into the training procedure. 

We did the experiments with 36-spin ($6\times 6$) triangular lattice with $J_2/J_1=1.3$ that corresponds to so-called quantum spin liquid phase with system being unable to acquire long-range magnetic order even at zero temperature \cite{PhysRevB.74.014408}. In absence of a well-defined magnetic order with some characteristic scale, quantum spin liquids can be regarded as critical (scale-invariant) systems, and require multi-scale methods of analysis. Moreover, in this regime, the ground state wave function has a complicated sign structure and hence represents a nice testbed for comparing different NQS architectures.

\begin{figure}[hbt]
\caption{Results of experiments with NQS representing sign structures of ground state of the quantum Heisenberg model on $6\times 6$ triangular lattice with periodic boundary conditions. We run each training procedure 10 times, and the average performance is shown. The shaded regions are $95\%$ confidence bands.}
\label{fig:nqs-triangular-6x6}
\centering
\includegraphics[width=80mm]{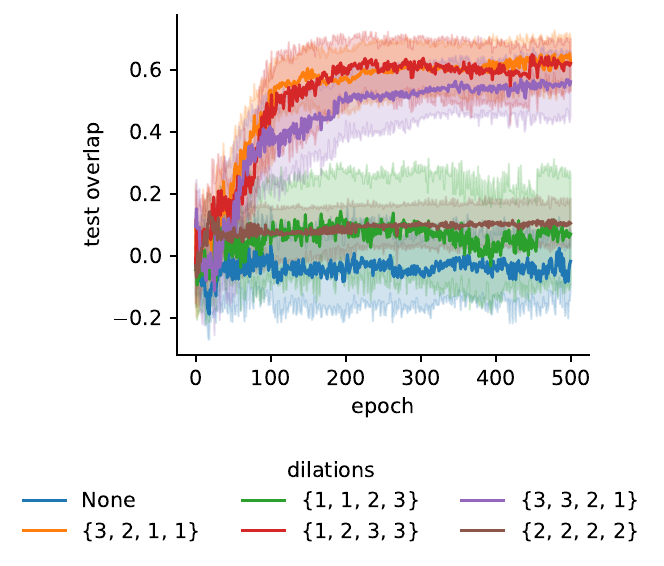}

\end{figure}

In Fig.~\ref{fig:nqs-triangular-6x6}, we depict the dependence of test overlap over the training epoch under the following settings. The train set is sampled from $p(i)$ with $\varepsilon_{train}=0.001$. Optimization is performed with Adam algorithm~\cite{kingma2017adam} at learning rate $10^{-3}$. The network has four layers, 32 channels at each layer, convolutional kernel size is 3, and various dilation schemes (i.e. distribution of dilations over the layers) are tried. We see that all networks with dilations perform better than the network without dilations. However, different dilation schemes can yield to rather different performance gains. Specifically, ``flat'' dilation scheme and dilation scheme without dilations at the first two layers perform just a little bit better than non-dilated network. Contrary to that, dilation schemes with various dilations make the model significantly better, increasing the test overlap from $\tilde{\cal O}<0.1$ to $\tilde{\cal O}\simeq 0.6$.

\begin{figure*}[hbt]
\caption{The performance of the neural networks of depth $3$ (top row) and $4$ (bottom row) for various values of $\varepsilon_{train}$. Kernel size is $3$. Results after 500 epochs of training are shown. Error bars are 95\% confidence intervals.}\label{fig:nqs-triangular-other}
\centering
\includegraphics[width=0.9\textwidth]{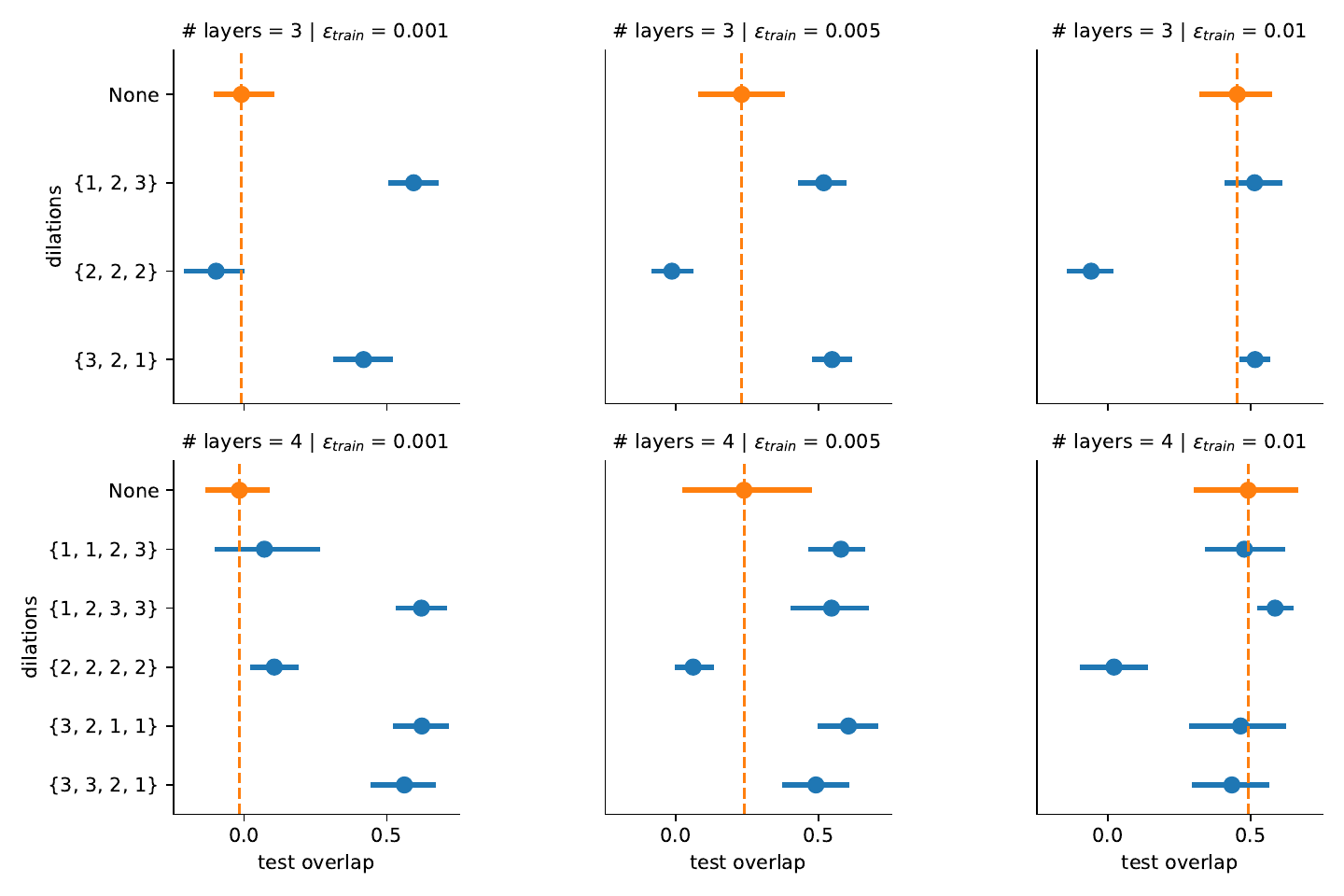}
\end{figure*}

We obtained similar results with various other settings, including choice of different $\varepsilon_{train}$, different depth of the neural network (three layers instead of two), different size of the kernel (two instead of three), see Fig.~\ref{fig:nqs-triangular-other} and Appendix~\ref{sec:additional-nqs} for details.
Our results seem to be rather robust with respect to those changes: in every settings we see that specific dilation schemes (usually like $\{3, 2, 1\}$ or $\{1, 2, 3\}$) work better than the absence of the dilations. As $\varepsilon_{train}$ is increased, the difference in learning quality between non-dilated networks and diverse dilation schemes is washed out. Given that the Hilbert space dimension scales exponentially with the number of quantum degrees of freedom, in realistic considerations of larger scale systems $\epsilon_{train} \rightarrow 0$, and the dilated networks can provide a strong advantage.


\section{Conclusions}
For scientific ML applications, usage of invariant architectures, in particular translationally invariant architectures, is a known receipt to overcome data scarcity, not involving data augmentation (which is another approach fulfilling this goal, but requiring additional computational resources for training). 
Several features such as absence of downsampling blocks distinguish translationally invariant CNN architectures from the traditional ones. Here we showed that by combining translational invariance with stacks of dilated convolutions at different scales one can considerably improve accuracy of such networks in the context of physical applications (photonics and quantum many body physics), where multi-scale nature of the systems of interest is important. 

These case studies can be viewed as a motivation to extend this approach onto more complex physical problems and more sophisticated neural network architectures. The dilations can be easily added to other types of networks (e.g Transformers \cite{DilTrans}), the combination of translational invariance + stacks of dilations is more complicated but worthwhile improvement.  
In the context of NQS, the specific task of learning sign structures can be integrated into complete procedure of obtaining variational neural approximations of target many-body quantum states, where signs and amplitudes are learned in an unsupervised manner.
In \cite{sharirVQ} a vector-quantized (VQ) transformer  architecture for neural quantum states (VQ-NQS) is discussed for this purpose. Our current results suggest that one of the promissing directions on this route could be a combination of VQ-transformer, translational invariance, and stacks of dilations.


\begin{acknowledgments}
This work used the Dutch national e-infrastructure with the support of the SURF Cooperative using grant no. EINF-7473. A.I. was partly supported by German Research Foundation (DFG) via SFB 986 ``Tailor-made Multi-scale Materials Systems: M3'', Project 192346071 and is thankful to A.~Petrov, M.~Eich, V.~Lempitsky for useful discussions. I.S. was supported by the Dutch Research Council (NWO) via the Spinoza Prize of M.~I.~Katsnelson.

\end{acknowledgments}

\appendix
\section{Additional experiments with neural quantum states}\label{sec:additional-nqs}
\begin{figure*}[hbt]
\caption{The performance of neural networks of depth $3$ (top row) and $4$ (bottom row) for various values of $\varepsilon_{train}$. Kernel size is $3$.}\label{fig:dep-3-4}
\centering
\includegraphics[width=0.9\textwidth]{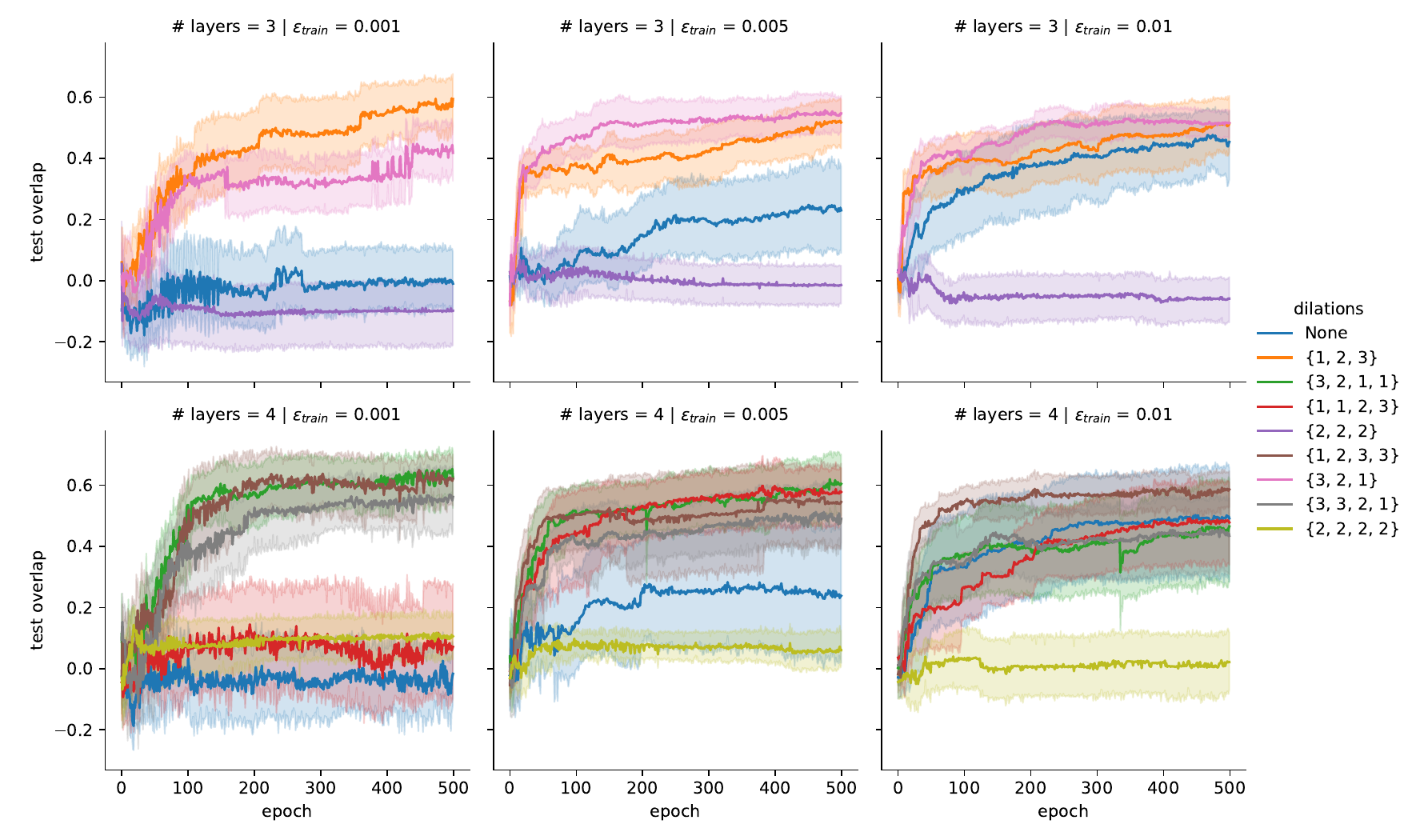}
\end{figure*}

\begin{figure*}
\caption{The performance of the neural networks of depth $3$, kernel size $2$.}\label{fig:depth-3-kernel-2}
\includegraphics[width=0.9\textwidth]{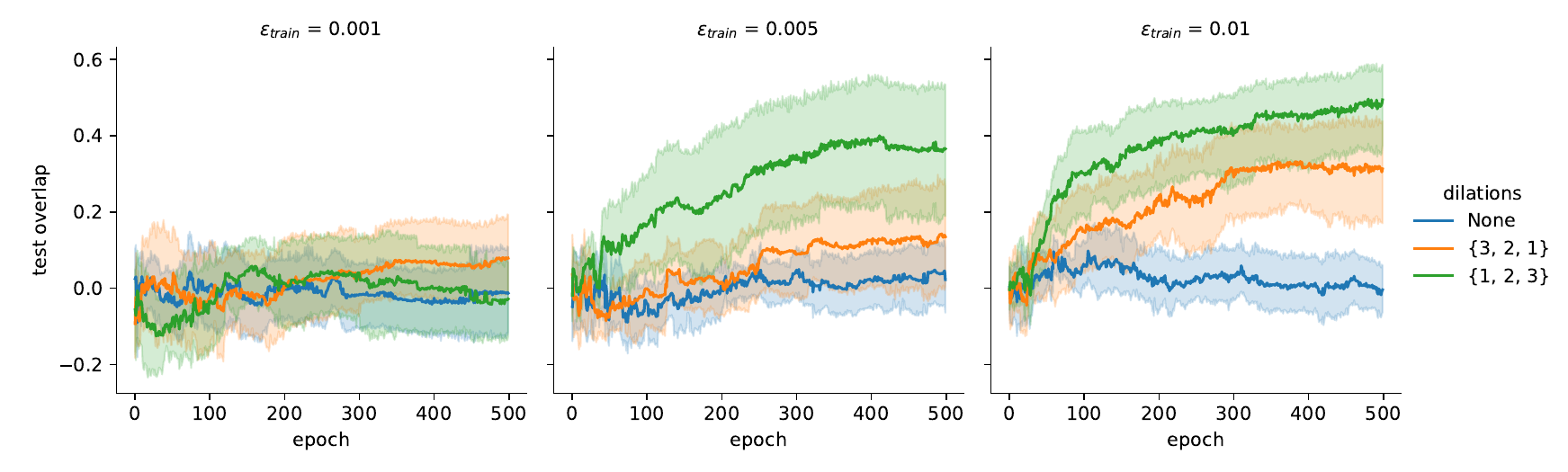}
\end{figure*}

Here we present the results of additional experiments with neural quantum states that demonstrate robustness of our conclusions, see Fig.~\ref{fig:dep-3-4} and Fig.~\ref{fig:depth-3-kernel-2}. The quantum system of interest is always $J_2/J_1=1.3$ Heisenberg antiferromagnet on $6\times 6$ periodic triangular lattice. The performance averaged over 10 runs is shown. The shaded regions are $95\%$ confidence bands. The neural network has 32 channels at each convolutional layer.


\clearpage

\section{Photonic datasets description and additional experiments with them}\label{sec:additional-PhC}

We used three datasets of photonic crystals for experiments with network architectures. The first one, $D_1$, was studied in \cite{MIT1} and contains binary images of $64\times 64$ pixel size, as described in the main text, and solutions of Maxwell eigenvalue equations for TM and TE modes on a $23 \times 23$ grid in reciprocal space. For this dataset, 6 bands were calculated, and the task for the network was to determine maxima and minima of each band.
The second ($D_2$) and the third ($D_3$) datasets were created following the procedure being described in \cite{MIT2}.
The $D_2$ dataset has 2000 images of photonic crystal unit cells of resolution $64\times 64$, and the third dataset 1000 images of unit cells with resolution $96 \times 96$. For both datasets, 17 first bands are calculated.
Most importantly, complexity of the unit cells is varied in controllable manner.
Following \cite{MIT2}, we parameterize $\epsilon(r)$ in each unit cell  by choosing a level set of a Fourier sum function $\Phi$, being defined as a linear sum of plane waves with frequencies evenly spaced in the reciprocal space (up to a certain cut-off $F$, which corresponds to the ``complexity'' of the unit cell):
\begin{equation}
    \Phi({\mbox r}) = \mbox{Re} \Bigl[ \sum_{n_x=-F}^{F} \sum_{ n_y=-F}^{F} c_{n_x,n_y} \mbox{exp} (2 \pi i (n_xx+n_yy)) \Bigr],
\end{equation}
where each $c_{n_x,n_y}$ is a complex coefficient of the form $r e^{i\theta}$, with $r=\sqrt{x^2+y^2}$, and $r, \theta/2 \pi$ being  separately sampled uniformly in [0, 1). Also, a filling fraction $\rho$ (defined as fraction of area of the unit cell filled with the material with permittivity $\epsilon_1$) is sampled uniformly in [0,1). The sampled $\rho$ defines the level $\Delta$ which determines binarisation of the sum function and therefore material distribution:
\begin{equation} 
\epsilon(\bf{r}) =   \left\{
\begin{array}{ll}
      \epsilon_1, & \Phi(\bf{r}) \leq \Delta \\
      \epsilon_2, & \Phi(\bf{r}) > \Delta \\
\end{array} 
\right.
\end{equation}
($\rho$ nonlinearly depends on $\Delta$, so if we would have sampled  $\Delta$ uniformly, we would obtain some complicated distribution of $\rho$).
With the MIT Photonic Bands (MPB) software \cite{mpb}, and freely available codes of \cite{MIT2}, we computed the band structure of each unit cell up to the first 17 bands, with the resolution of  $13\times 25$ over half of the Brillouin zone $ 0 \leq k_x,k_y \leq \pi$. We slightly modified the original codes of \cite{MIT2}, where k-grid was $25\times 25$ in the interior of the Brillouin zone $ -\pi < k_x,k_y < \pi$. Half of the Brillouin zone is enough for finding the maxima and minima due to central symmetry, while including boundary of the zone provide better accuracy of the extrema determination. 
The $D_2$ dataset has unit cells of resolution $64\times 64$ and maximal "complexity" cutoff factor $F=4$ (to be more precise, there are 500 images each having one of the  F=1,2,3,4 factors, in total 2000 images),  and the third dataset has 1000 images in total and factor $F=8$ (with 125 images of each of the factors F from 1 to 8).


\begin{table*}[t]
\begin{tabular}{ |p{3cm}||p{3cm}|p{3cm}|p{3cm}|p{3cm}| }
 \hline
 \multicolumn{5}{|c|}{ Optuna experiments with $D_1$ dataset, $N_{train}$=1600} \\
 \hline
 Parameter & Session 1  & Session 2 & Session 3 & Session 4 \\
 \hline
 Channels sequence   & $\{1,1,1,1,1\}$    & $\{1,1,1,1,1\}$  & $\{1,1,2,2,4\}$ & $\{1,1,2,2,4\}$   \\
Stride sequence   & $\{1,1,1,1,1\}$    & $\{1,1,1,1,1\}$  & $\{1,1,1,1,1\}$ & $\{1,2,1,2,1\}$   \\
 
 Variants of dilation orderings&   $[V_1,V_2,V_3,V4]$  & $[V_4] $ & $[V_4] $ & $[V_4] $ \\
 Variants of CFactor value &   $[24,48]$  & $[24,48] $ & $[12,24] $ & $[12,24]$ \\
 Nr of trials & 64 & 32 & 32 & 32 \\
 Nr of variants in the configuration space  & 128 & 32 & 32 & 32\\
  \hline
    \multicolumn{5}{|c|}{Results of the best trial}  \\
    \hline 
  Nr. of parameters &  88.5k  & 81k & 95.6k & 95.6  \\ 
 CFactor&   48  & 48 & 24 & 24  \\
 Dilation variant & $V_1$  &  -  &  -  &  -   \\
 Filter size, 1st layer & 4  &  4   &  2  &  1 \\
 Filter size, last layer & 2  & 3 &  1   &  3  \\
 MAPE ``on the best epoch'' & {\bf 0.51}  &  0.72  & 0.72  &  0.71     \\
 \hline
\end{tabular}
 \caption{Optuna experiments with the $D_1$ dataset. Number of training samples: 1600, number of convolutional layers: 5, number of FC layers: 3. 4 variants of ordering dilations were considered: $V_1: \{16,8,4,2,1\} $, $ V_2: \{1,2,4,8,16\} $,$ V_3: \{1,2,4,8,12\} $, $ V_4: \{1,1,1,1,1\}$,. In the variant $V_4$, no dilations were used. In Session 1, configurational space contained dilation ordering variants $V_1,V_2,V_3,V_4$ while in Sessions 2-4 the dilation variant was fixed to be $V_4$. 
 In Session 4, stride sequence $  \{1,2,1,2,1\}$ was used, whereas in Sessions 1-3 it was fixed to be $ \{1,1,1,1,1\}$.
 $\mathrm{CFactor}$ was chosen from the set $[24,48]$ (except in Sessions 3-4, where sequence of channels numbers was non-uniform, and $\mathrm{CFactor}$ was chosen from the set $[12,24]$ ). The number of channels in the convolutional part were either growing like $\mathrm{CFactor} \times \{1,1,2,2,4\}$  (Sessions 3,4), or remaining constant (Sessions 1,2). From 32 to 64 trials were done in every session. Bottom panel shows results of the best trials in each session.}
  \label{tab:D1.0}
\end{table*}

\begin{table*}[t]
\begin{tabular}{ |p{3cm}||p{3cm}|p{3cm}|p{3cm}|p{3cm}| }
 \hline
 \multicolumn{5}{|c|}{ Optuna experiments with $D_1$ dataset, $N_{train}=800 $ } \\
 \hline
 Parameter & Session 1  & Session 2 & Session 3 & Session 4 \\
 \hline
 Channels sequence   & $\{1,1,1,1,1\}$    & $\{1,1,1,1,1\}$  & $\{1,1,2,2,4\}$ & $\{1,1,2,2,4\}$   \\
Stride sequence   & $\{1,1,1,1,1\}$    & $\{1,1,1,1,1\}$  & $\{1,1,1,1,1\}$ & $\{1,2,1,2,1\}$   \\
 
 Variants of dilation orderings&   $[V_1,V_2,V_3]$  & $[V_3] $ & $[V_3] $ & $[V_3] $ \\
  Variants of FCN configurations&   $[F_1,F_2]$  & $[F_1,F_2] $ & $[F_1,F_2] $ & $[F_1,F_2] $ \\
 Nr of trials & 48 & 32 & 32 & 32 \\
 Nr of variants in the configuration space  & 96 & 32 & 32 & 32\\
  \hline
    \multicolumn{5}{|c|}{Results of the best trial}  \\
    \hline 
  Nr. of parameters &  99.4k  & 99.7k & 72.7k & 72.8k  \\ 
 CFactor&   48  & 48 & 24 & 24  \\
 Dilation variant & $V_1$  &  -  &  -  &  -   \\
 Filter size, 1st layer & 2  &  3   &  2  &  3 \\
 Filter size, last layer & 3  & 3 &  2   &  2  \\
 FC variant & $F_1$  & $F_1$ &  $F_1$   & $F_1$ \\
 MAPE ``on the best epoch'' & {\bf 0.59}  &  0.80  & 0.78  &  0.77     \\
 \hline
\end{tabular}
 \caption{Optuna experiments with the $D_1$ dataset. Number of training samples: 800, number of convolutional layers: 5, number of FC layers: 3. 3 variants of ordering dilations were considered: $V_1: \{16,8,4,2,1\} $, $ V_2: \{1,2,4,8,16\} $,$ V_3: \{1,1,1,1,1\}$. In the variant $V_3$, no dilations were used. In Session 1, configurational space contained dilation ordering variants $V_1,V_2,V_3$ while in Sessions 2-4 the dilation variant was fixed to be $V_3$. 
 Configuration of FC layers were chosen from 2 variants: $F_1: \{32,100,100\} $, $ F_2: \{24,120,100\} $. The number of channels in the convolutional part were either growing like $\mathrm{CFactor} \times \{1,1,2,2,4\}$  (Sessions 3,4), or remaining constant (Sessions 1,2). In the latter case, CFactor was fixed to be 48, in the former it was fixed to 24. From 32 to 48 trials were done in every session. Bottom panel shows results of the best trials in each session.}
  \label{tab:D1.01}
\end{table*}

\begin{figure}
\caption{(a) Distribution of bandgaps between the first and the second band in 1000 samples of the third dataset.
(b) Distribution of all bandgaps in the third dataset. We evaluate bandgaps between all adjacent bands in every sample of the dataset (16 bandgaps for each sample, between kth and (k+1)-th bands for k=1,..16). Histogram of the obtained in such way 16000 numbers is shown on the figure.}
\label{FigH}
\centering
\includegraphics[width=65mm]{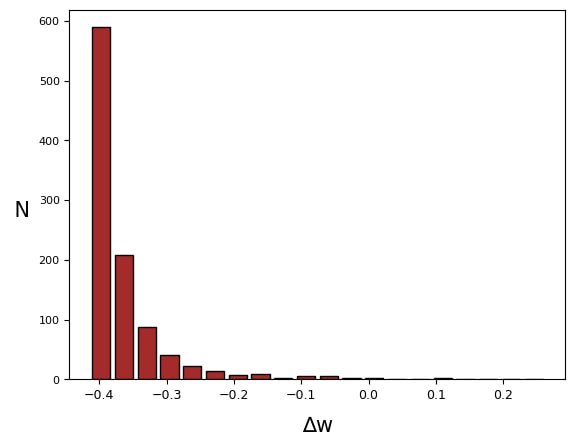}
\includegraphics[width=65mm]{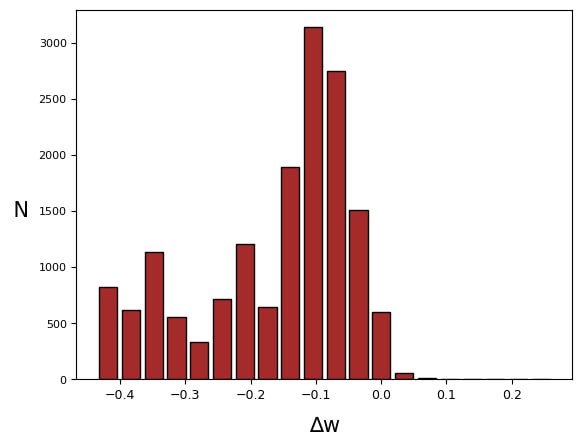}
\end{figure}

Experiments with the third dataset (1000 unit cells of resolution $96\times 96$) were shown in Tables~ \ref{tab:D3.2}, \ref{tab:D3.3} in the main text. Here we also show histogram of distribution of bandgaps for this dataset in Fig.~\ref{FigH}.
Experiments with dataset $D_1$ are shown in Tables~\ref{tab:D1.0}, \ref{tab:D1.01} and are discussed in the main text.

\clearpage

\nocite{*}

\bibliography{apssamp}

\end{document}